# Evaluation of (GPS/GLONASS) patch versus RF GPS (L1) Patch Antenna performance parameter

Gholam Aghashirin[1], Hoda S. Abdel-Aty-Zohdy[1], Mohamed A. Zohdy[1], Darrell Schmidt[2], and Adam Timmons[3]

[1]Department of Electrical and Computer Engineering, Oakland University, Rochester, Michigan, USA

aghashirin@oakland.edu, zohdyhsa@oakland.edu, zohdyma@oakland.edu

[2]Department of Mathematics and Statistics, Oakland University, Rochester, USA
schmidt@oakland.edu

[3]Department of Mechanical Engineering, McMaster University, Hamilton, Canada
adam.timmons@me.com

ABSTRACT

*In any wireless communication network and system an antenna is an important element along the pathway and/or propagation path of an electrical signals. An addition, antenna module is a vital component of automated driving systems, it should function as needed in dGPS, HD map correction services, and radio and navigation systems.*

*The main scope, objective and goal of this engineering research work involves the evaluation and determining the performance parameter and characteristic of the dual band (GPS/GLONASS) patch vs RF GPS L1(1.57542 GHz) passive patch antenna characteristic. FEKO simulation studies are carried out to extensively compare, make an assessment and evaluate the characteristic and performance parameter, such as the average gain and/or passive gain of the proposed antenna in the presence of background noise. Prior to the start of the FEKO simulation studies, a physical mechanical dimension measurements via a Digital instrumentation were conducted for the following:*

*Radiating Element Size: The actual length (L), and width (W)*

*Substrate Material Size: The substrate length (Lsub), width (Wsub), and height (h)*

*The proposed antenna model for GPS only patch antenna operating at 1.57542 GHz and the GPS/GLONASS patch antenna resonating at 1.5925 GHz are developed. To be specific, this work presents the design, modeling, determining passive gain of the RF GPS L1 passive patch vs. active GPS/GLONASS patch antenna with intended targeted applications within the automotive system and space. Simulation are undertaken to generate the RF GPS L1 passive patch and active GPS/GLONASS patch antenna structure respectively for the sole purpose of evaluating the performance of the proposed GPS/GLONASS antenna. Simulation are performed rather than mathematical modeling. The two antennas are also compared from the size standpoint. The goal of this paper is to test, measure and evaluate the performance of GPS against GPS/GLONASS antennas. Another emphasis of this paper is how to obtain the equivalent amount of total passive gain in a GPS vs. that of GPS/GLONASS antenna.*

## 1. INTRODUCTION

GPS constellation system that is managed and maintained by the United States Department of Defense since the early 70s. GPS is a space base radio navigation system that supplies an estimated position, velocity and timing information to a GPS antenna/receiver module on the global. GPS system is mainly consisting of three parts, namely the space component, ground control station and user segment. GPS constellation we have 31 fully operational satellites positioned at an altitude of 20, 000 km (12, 427 miles) and it moves around the earth twice a day. Where each satellite transmits ranging and navigation data on the L1 (1.57542 GHz), L2 (1.22760 GHz), and L5 (1.17645 GHz) carrier frequencies [3].

In order to support an array of applications across various industries, wireless communication (Satellite, Spacecraft), Missile, Mobile Devices such as a Cell Phone, Ground Reference Station, spaces, platforms and automotive sector that requires accurate timing and precious positioning data, messages, signals information to be deliver to a domestic and non-domestic vehicles destined to USA, Canada, Mexico and European markets plus as a result of rapid growth in an automotive applications related to connected services, automotive radio head unit features, functions and navigation systems, an antenna that can receive both RF GPS L1 passive patch and active GPS/GLONASS patch antenna frequencies is needed. Therefore, due to its low cost and profile targeting applications across various sectors of automotive and non-automotive industries, the following two proposed antennas is investigated and compared in this research work:

1) RF GPS L1 passive patch antenna

2) Active GPS/GLONASS patch antenna

The RF GPS L1 passive patch antenna has radiating element size of 12.25 x 12.25 mm and it operates at the GPS L1 1.57542 GHz frequency. While on the other hand, the active GPS/GLONASS patch antenna has 12.25 x 12.25 mm with resonate frequency of 1.5925 GHz.

The RF GPS L1 passive patch and active GPS/GLONASS patch is model, simulated, designed, simulation results obtained and constructed on a 5.5 mm ceramic/porcelain substrate material and ground plane length and width size for each antenna of 95 by 95 mm respectively in the CADFEKO page were defined and antenna far field/passive gain were obtained in the

POSTFEKO environment. To be more direct, FEKO simulation software package was used to model, simulate and design both antennas.

We are interested to know the antenna characteristic performance parameter, such as the passive gain of the two proposed antenna. Initially, both antennas design parameters were defined, FEKO models for each respective presented antenna were created, by creating the radiating element, dielectric substrate, ground plane, feedpin, add a port to the created feedpin, added a voltage source of excitation, set the simulation frequency, union antenna parts together, mesh the generated antenna model geometry followed by simulating the model by running the FEKO Solver all within the CADFEKO environment.

POSTFEKO was used to obtain the passive gain expected results, 2D plots of simulated Far Field, top/side view of the studied antennas and use the simulation results from the POSTFEKO environment to compare passive gain results and determine an optimal antenna solution and make recommendations for the automotive industries with intended applications targeting automotive radios and navigation systems that requires GPS and GPS/GLONASS signals.

Upon comparing the obtained simulation results of the RF GPS L1 patch and GPS/GLONASS patch antenna structure from the POSTFEKO environment in this research work, it was found and concluded that the simulated results of the proposed GPS/GLONASS antenna is a better antenna solution for the automotive applications related to automotive radios and navigation systems.

An antenna could be defined as a wireless communication device or module such as a piece of wire for radiating or receiving electromagnetic wave propagating in a communication channel, such as guided structure transmission line and then getting transmitted into a free space and/or vice versa in the receiving mode. Furthermore, we present the passive gain of the RF GPS only patch and GPS/GLONASS antenna structure using FEKO electromagnetic simulation software package, in order to support automotive applications. Plus, this study describes the modeling, design, simulation and analysis of RF GPS only (L1) patch and GPS/GLONASS patch antenna. According to Constantine A. Balanis, the antenna is the transitional structure between free-space and a guiding device, for wireless communication systems, the antenna is one of the most critical components. For the past few decade Microstrip Patch Antenna were used heavily in high performance aircraft, spacecraft, satellite and missile where size, weight, cost, performance, ease of installation, and aerodynamic profile are constraints. Low profile antennas maybe required [1] for packaging and/or aesthetic constraints. The active GPS/GLONASS patch antennas play a significant role in today's modern communications, i.e. they nicely meet automotive specification requirements, most antenna designers and OEMs mainly preferred and select this rectangular/square active GPS/GLONASS patch antenna, in order to mount, install, place and position it on their production vehicles.

This research work main contribution is the RF GPS L1 passive only patch versus active GPS/GLONASS passive gain comparison through modeling and simulation within the FEKO (CADFEKO and POSTFEKO) environment.

The question/problem to address of how to obtain the same and/or equivalent amount of antenna passive gain in a GPS/GLONASS versus GPS L1 antenna is the objective of this research work. The problem has not been solved, no solution is available on the shelf.

The verification and validation of antenna characteristic performance parameter testing of proposed antenna can be conducted at the automotive bench, vehicle level, and at an antenna range with no limitations or deficiencies by using antenna measurements with a presented Antenna Under Test (AUT) component, Antenna Measurement Software, DC Power Supply,

Bias-Tee, and Vector Network Analyzer (VNA). The accurate far field and/or proposed antenna passive gain measurements can be performed fast.

## 1.1. Patch Antenna Geometry Photo

Figure 1 below shows the patch antenna basic structure model in general, where we have a metallic rectangular radiating element, which we call it patch that is place over a dielectric substrate material with a specific relative permittivity and at the bottom of substrate is a conducting base layer that acts as a rectangular conducting ground plane.

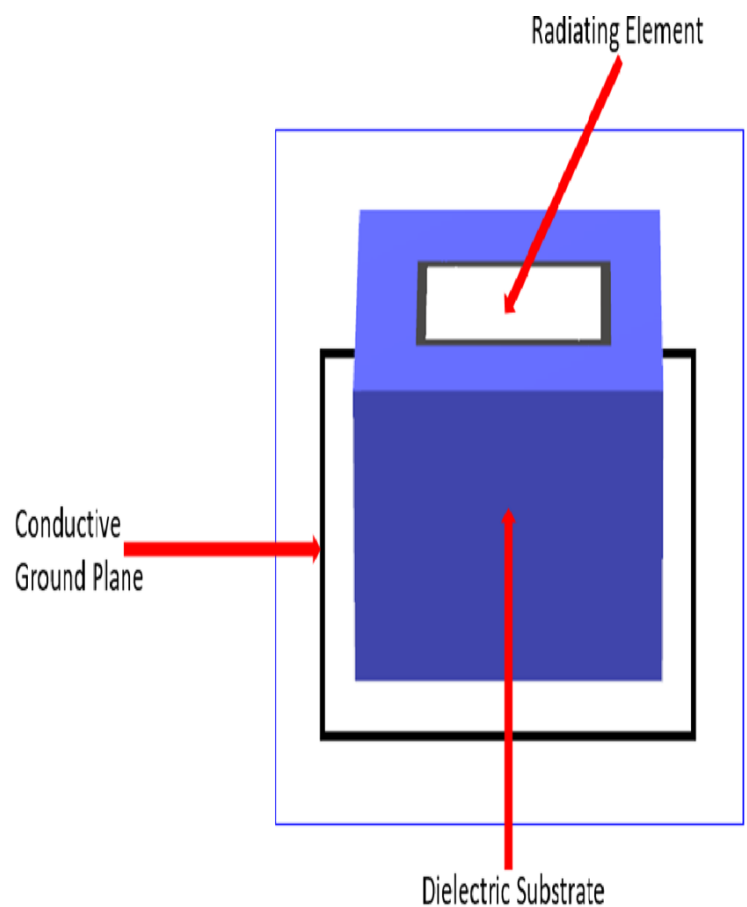

Figure 1. Patch Antenna Structure

## 1.2. GPS Constellation System Photo

Figure 2 displays the GPS Satellites in six orbital planes in space

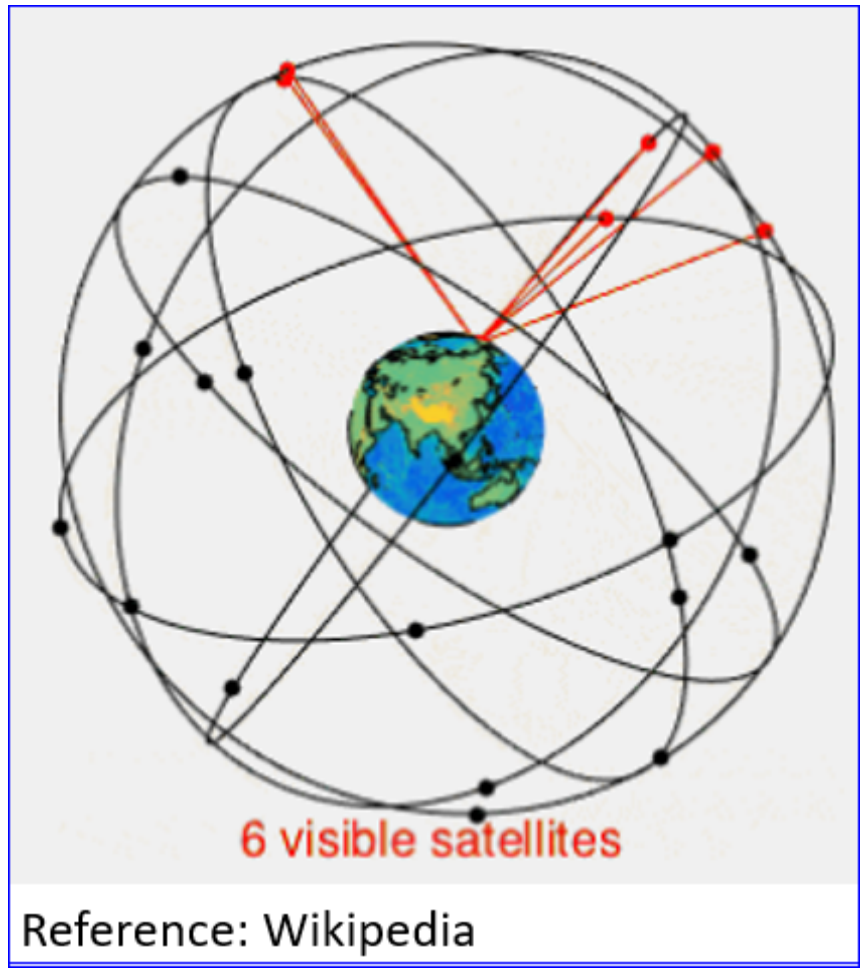

Figure 2. GPS Satellites Constellation

This research work is organized into six Sections. In Section 1, we highlight our Introduction followed by a Patch Antenna skeleton geometry in general. In Section 2, we are discussing Antenna Performance Characteristics and set of Design Equations. In Section 3, we offer Antenna Design, FEKO Simulation Studies, and Discussion of Results. In Section 4, we demonstrate FEKO Design Parameters and further Simulation Results. In Section 5, we present our Findings and Conclusions. Lastly, in Section 6, we have our Future Work Suggestions.

## 2. ANTENNA PERFORMANCE CHARACTERISTICS

In the following below are illustration of mathematical model/equations that define the antenna gain, efficiency, directivity, input impedance, patch actual length and width dimension.

**Equation 1** Expresses, the Antenna Gain as:

$$gain = 4\pi \frac{radiation\ intensity}{total\ input(accepted)power} = 4\pi \frac{U(\theta,\varphi)}{P_{in}} \quad (dimensionless) \qquad [1]$$

Also, Antenna Gain is defined in terms of Antenna Efficiency and Antenna Directivity according to Fawwaz T. Ulaby:

$$G = \varepsilon D \quad \text{(dimensionless)} \tag{2}$$

Where G=Antenna Gain, ε=Antenna Efficiency, and D=Antenna Directivity

**Equation 2** Expresses the Antenna Radiation efficiency as follows:

$$\varepsilon = \frac{Prad}{Pt} \quad \text{(dimensionless)} \tag{2}$$

Where Prad = Radiated Power, Pt = Transmitter Power

In general, the overall Antenna Efficiency can be express as below

$$e_0 = e_r\, e_c\, e_d \tag{1}$$

Where $e_O$ = Total Efficiency     (dimensionless)

$e_r$ = reflection (mismatch) efficiency = $(1 - |\Gamma|^2)$     (dimensionless)

$e_c$ = Conduction Efficiency     (dimensionless)

$\Gamma$ = Voltage reflection coefficient at the input terminals of the antenna

$e_d$ = Dielectric Efficiency     (dimensionless)

**Equation 3** Expresses, the Antenna Directivity

$$D = 4\frac{\pi}{\Omega_P} \; (dimensionless) \tag{2}$$

Where $\Omega_P$ = Pattern Solid Angle = $\iint_{4\pi} F(\theta, \varphi)(d\,\Omega)$ and

$F(\theta, \varphi)$ = Normalized Radiation Intensity = (Elevation Angle, Azimuth Angle)

**Equation 4** Highlights, the Antenna Input Impedance, defined as:

Input Impedance = $Z_A$ = $R_A$ + $jX_A$     (ohms)     [1]

Where $Z_A$ = antenna impedance at the input terminals of an antenna when it operates in transmitting mode (ohms)

$R_A$ = antenna resistance at the input terminals of an antenna when it operates in transmitting mode (ohms)

$X_A$ = antenna reactance at the input terminals of an antenna when it operates in transmitting mode (ohms)

In general, the $R_A$ parameter from below **Equation** is mainly made up of two resistances ($R_r$ and $R_L$) of the antenna

Resistive component = $R_A = R_r + R_L$     (ohms)     [1]

Where

$R_r$ = Represents the radiation resistance of the antenna    (ohms)

$R_L$ = Represents the loss resistance of the antenna    (ohms)

If we assume that the antenna is connected/attached to a signal/function generator/source with internal impedance, when the antenna is used in the transmitting mode of operation then internal impedance is defines as listed below:

Internal impedance $(Z_g) = R_g + jX_g$     (ohms)     [1]

Where

$R_g$ = Represents the resistance of signal source/generator impedance    (ohms)

$X_g$ = Represents the reactance of signal source/generator impedance    (ohms)

Solving these equations above at high level will permits to obtain the some of the antenna characteristics. Where on the other hand FEKO simulator/simulation software package is based on the Method of Moments (MoM) integral formulation of James Maxwell's equations, in order to solve for antenna characteristics, such as antenna gain, antenna input impedance, etc.

Per Constantine A. Balanis the following simplified rectangular patch antenna equations and formulas can be utilized to design microstrip antenna for a given relative permittivity and/or dielectric constant of the substrate material (Epsilon r, $\varepsilon r$), the resonate frequency (fr), and the substrate height (h):

**Equation 5** Determine the patch width W:

Patch Width = W = $\frac{v0}{2fr}\sqrt{2/(\varepsilon r + 1)}$     (cm and/or in)     [1]

Where v0 = The velocity of light in free-space (Constant value)

fr = The resonant frequency

$\varepsilon r$ = The dielectric constant of the substrate

**Equation 6** Determine the actual patch length L:

Patch Length = L = $\frac{1}{2fr\sqrt{\varepsilon reff}\sqrt{\varepsilon 0 \mu 0}} - 2\Delta L$     (cm and/or in)     [1]

Where $\Delta L$ = The extended incremental length of the patch =

h $(0.412) \times (\varepsilon reff + 0.3)\left(\frac{W}{h} + 0.264\right) / (\varepsilon reff - 0.258)\left(\frac{W}{h} + 0.8\right)$

fr = The resonant frequency

$\varepsilon reff$ = The effective dielectric constant = $\frac{\varepsilon r + 1}{2} + \frac{\varepsilon r - 1}{2} \times \left(1 + 12\frac{h}{W}\right)^{-1/2}$

$\varepsilon 0$ = The permittivity of free space

$\mu 0$ = The permeability of free space

h = The substrate height

W = The patch width

$\varepsilon r$ = The dielectric constant of the substrate

Equation 5 and equation 6 above allows for the computation of a rectangular microstrip antenna mechanical dimensions such the actual patch length and width for the purpose of a practical design of patch antenna.

## 3. ANTENNA DESIGN, ASSESSMENT, SIMULATION STUDIES AND DISCUSSION OF RESULTS

The testing, experimental, comparison and evaluation of the square RF GPS only passive patch and active GPS/GLONASS patch antenna design and simulation of the proposed two antennas is performed using FEKO which has not been previously investigated and/or studied at the FEKO simulation level. The RF GPS only passive patch and active GPS/GLONASS patch antenna will be compared and contrasted mostly from the total passive gain in FEKO simulation environment viewpoint. The photo of each of the respective two test and presented antennas under evaluation and assessment are outlined and shown in Figure 3, Figure 4, Figure 5, and Figure 6. Figure 3 depicts the Front View of the dual band active GPS/GLONASS antenna. From the Figure 4 we can see the Top View of the dual band active GPS/GLONASS antenna. Figure 5 depicts the Front View of the RF GPS only L1 square/rectangular passive patch antenna. From the Figure 6 we can see the Top View of the RF GPS only L1 square/rectangular passive patch antenna.

### 3.1. Active GPS/GLONASS Patch Antenna Front View Photo

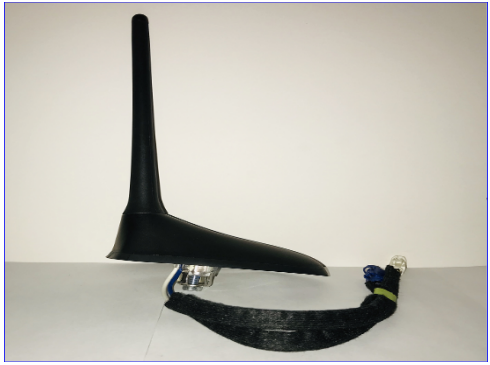

Figure 3. Dual Band Constellation Active GPS/GLONASS Patch Antenna

## 3.2. Active GPS/GLONASS PatchAntenna Top View Photo

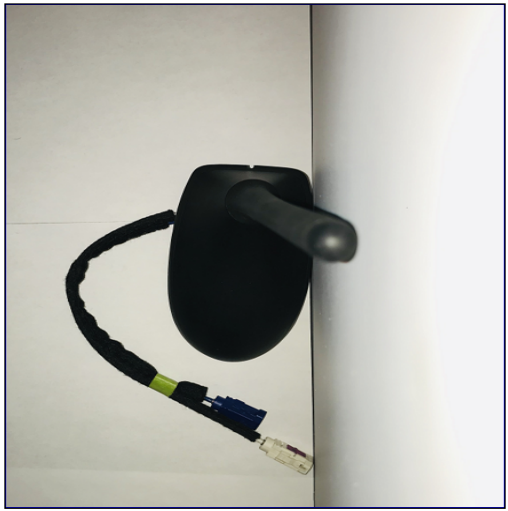

Figure 4. Dual Band Constellation Active GPS/GLONASS Patch Antenna

## 3.3. RF GPS Only Passive Patch Antenna Front View Photo

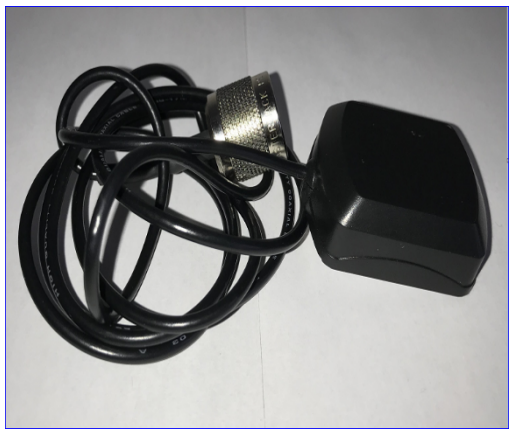

Figure 5. RF GPS Only Passive Patch Antenna

## 3.4. RF GPS Only Passive Patch Antenna Top View Photo

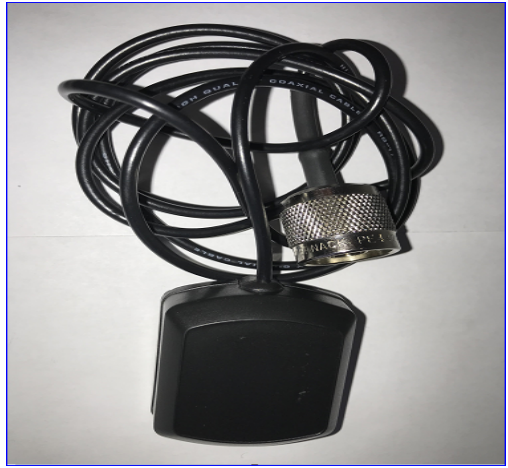

Figure 6. RF GPS Only Passive Patch Antenna

### 3.5. RF GPS Antenna Samples Consideration for the Evaluation

Table 1. One sample of each antenna used for the model simulation in the FEKO environment

| Dual-Band Active GPS/GLONASS Patch Antenna | 1 |
|---|---|
| RF GPS Only Passive Patch Antenna | 1 |

### 3.6. Range of Operating and Simulation Frequency Requirements

The active GPS/GLONASS patch, RF GPS Only L1 passive patch antenna frequencies were used for the purpose of this paper test and simulation activities:
- RF GPS Only L1 passive patch antenna Frequency, GPS (L1): 1.57542 GHz
- Active GPS/GLONASS patch antenna: 1575 to 1610 MHz, Center Frequency (fc)= 1.5925 GHz)

The proposed antennas [RF GPS L1 Passive Patch and active GPS/GLONASS] patch antenna characteristic was simulated by using FEKO simulation software package. An analysis was conducted next and finally total gain for each of the sample antenna were observed from the POSTFEKO environment.

### 4. FEKO DESIGN PARAMETERS AND SIMULATION RESULTS

We initially performed and conducted the proposed two antennas physical dimension measurements of patch length/width, substrate length/width/height followed that by antenna modelling, simulation and data results analysis activities within the CADFEKO and POSTFEKO software package domain for the following antenna samples:
- RF GPS Only L1 passive patch antenna
- Active GPS/GLONASS patch antenna

### 4.1. Antenna Samples for FEKO Simulation

Table 2. Antenna substrate and radiating element dimensions

| Patch Size & Application | Reference & Device Under Test (DUT) Antenna |
|---|---|
| RF GPS L1 Only Single Passive Patch Antenna (L1-, 1.57542 GHz)<br>• Substrate Size: 24.9 x 24.8 x 4.5 mm<br>• Radiating Element Size: 12.25 x 12.25 mm | Reference |
| Dual Band Active GPS/GLONASS Antenna (1.5925 GHz)<br>• Substrate Size: 24.7 x 24.7 x 4.5 mm<br>• Radiating Element Size: 12.25 x 12.25 mm | DUT |

## 4.2. Design Parameters within FEKO Simulation Environment

*Table 3.* FEKO Mesh and Loss Tangent Parameters for RF GPS only Passive Patch and Active GPS/GLONASS Patch Antenna

| RF GPS and GPS/GLONASS Antenna Component | Parameter | Value |
|---|---|---|
| RF GPS only Passive Patch Antenna Operating at , 1.57542 GHz | Mesh-Wire Segment Radius | 1.587e-3 mm |
| | Dielectric Loss Tangent for Porcelain Material | 2.1e-14 mm |
| Active GPS/GLONASS Patch Antenna Operating at 1.5925 GHz | Mesh-Wire Segment Radius | 1.569e-3 mm |
| | Dielectric Loss Tangent for Porcelain Material | 2.0e-14 mm |

## 4.3. Reference RF GPS Only L1 Passive Patch Antenna

Modeling, design, and simulation based on the following design parameters listed in Table 4 below.

*Table 4.* Design Parameters of Reference RF GPS only Passive Patch Antenna Structure

| Parameter | Value |
|---|---|
| Feed Length | 0.5 mm |
| Operating Frequency | 1.57542 GHz |
| Ground Plane Length | 95 mm |
| Ground Plane Width | 95 mm |
| Radiating Element Length | 12.25 mm |
| Radiating Element Width | 12.25 mm |
| Substrate Length | 24.8 mm |
| Substrate Width | 24.9 mm |
| Substrate Thickness | 4.5 mm |

| Substrate Dielectric Constant (Relative Permittivity) for Ceramic/Porcelain Material | 5.5 mm |
| --- | --- |

Figure 7 outlined below indicates the CADFEKO RF GPS patch antenna model that is pin fed voltage source of excitation (1 V, 50 Ω) Top View, Figure 8 plot below shows graphic representation of the Cross Section Image of the GPS only (L1 frequency, 1.57542 GHz) passive patch antenna on a finite square/rectangular orange color ground plane, square purple color substrate with dielectric constant value of 5.5 mm and a square dark blue color radiating element created in POSTFEKO. Figure 9 and Figure 10 depict the Front and Side View of the RF GPS only passive patch antenna generated in the POSTFEKO environment.

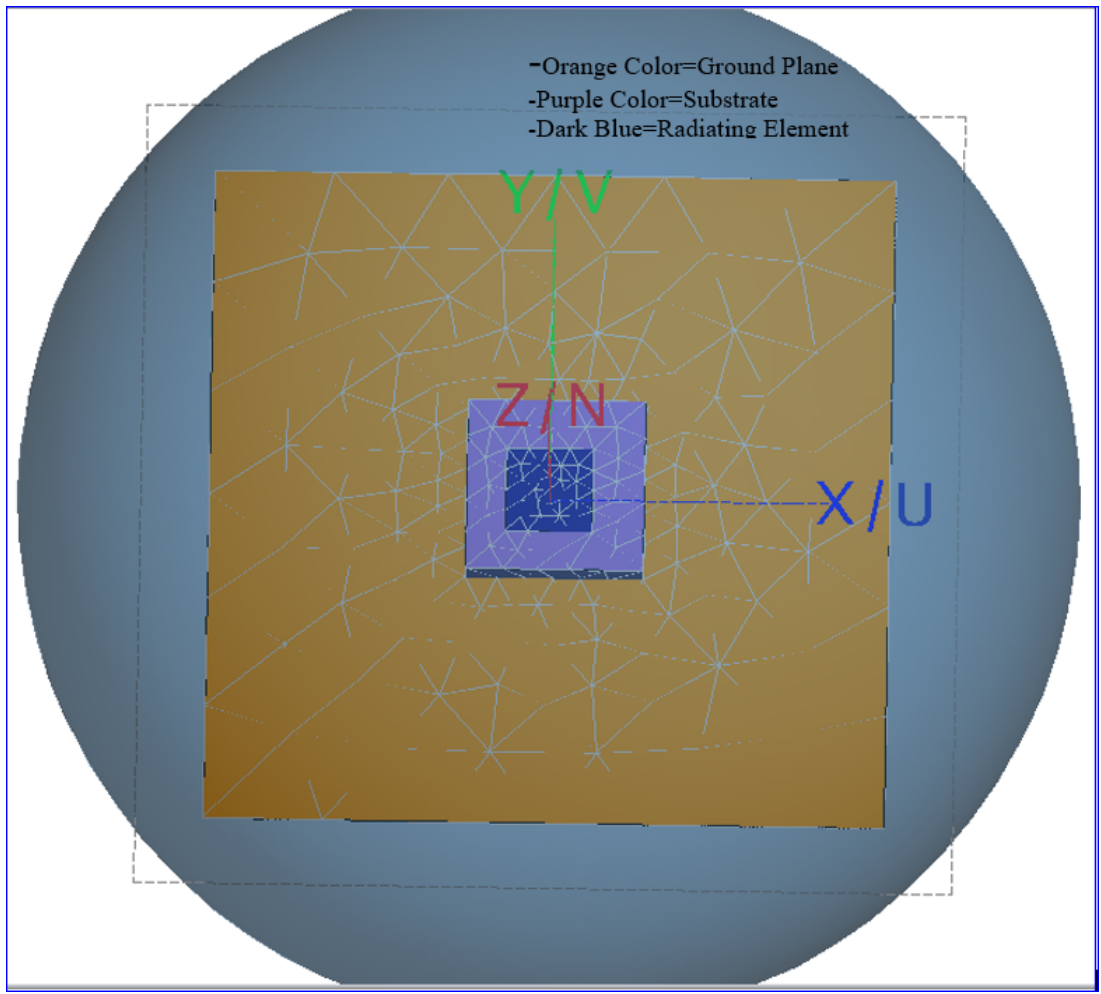

Figure 7. RF GPS Patch antenna operating at 1.57542 GHz (Top View-CADFEKO v2020)

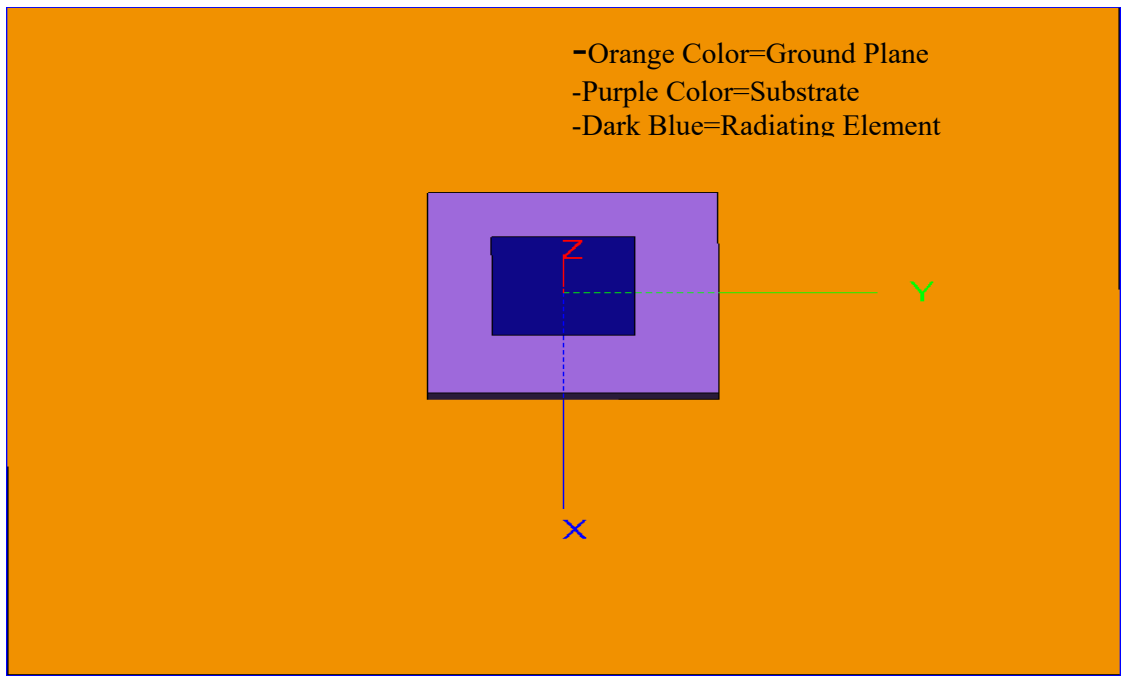

Figure 8. RF GPS Patch operating at 1.57542 GHz (Cross Section Image-POSTFEKO v2020)

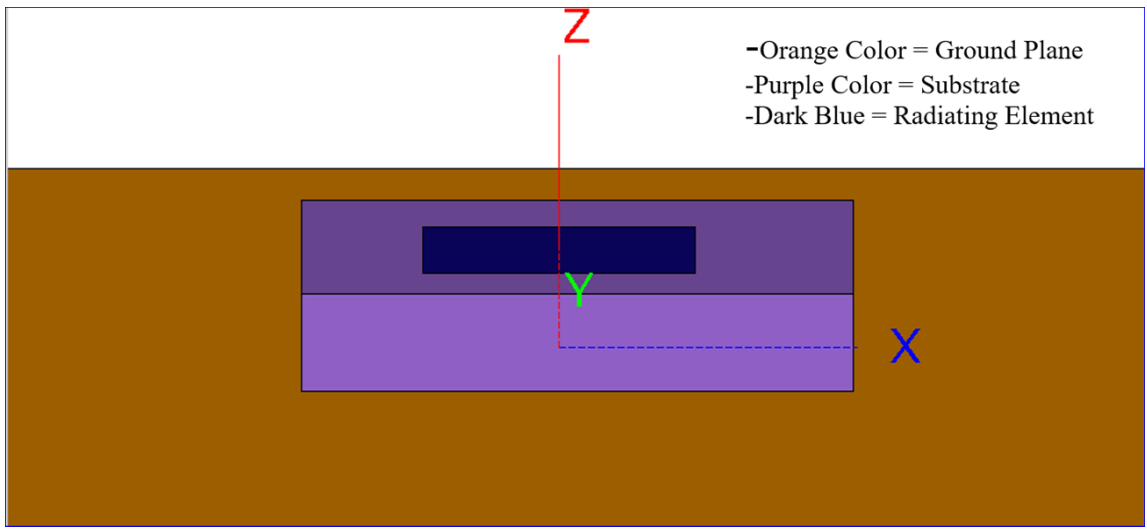

Figure 9. RF GPS Passive Patch operating at 1.57542 GHz (Front View- POSTFEKO v2020)

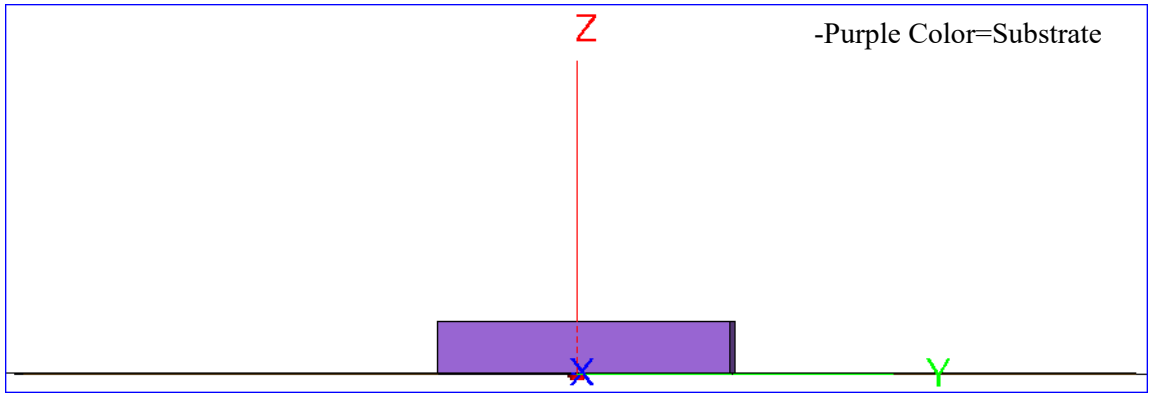

Figure 10. RF GPS Passive Patch operating at 1.57542 GHz (Side View- POSTFEKO v2020)

**4.4. Simulated Far Field of Reference RF GPS only Structure Passive Patch Antenna**

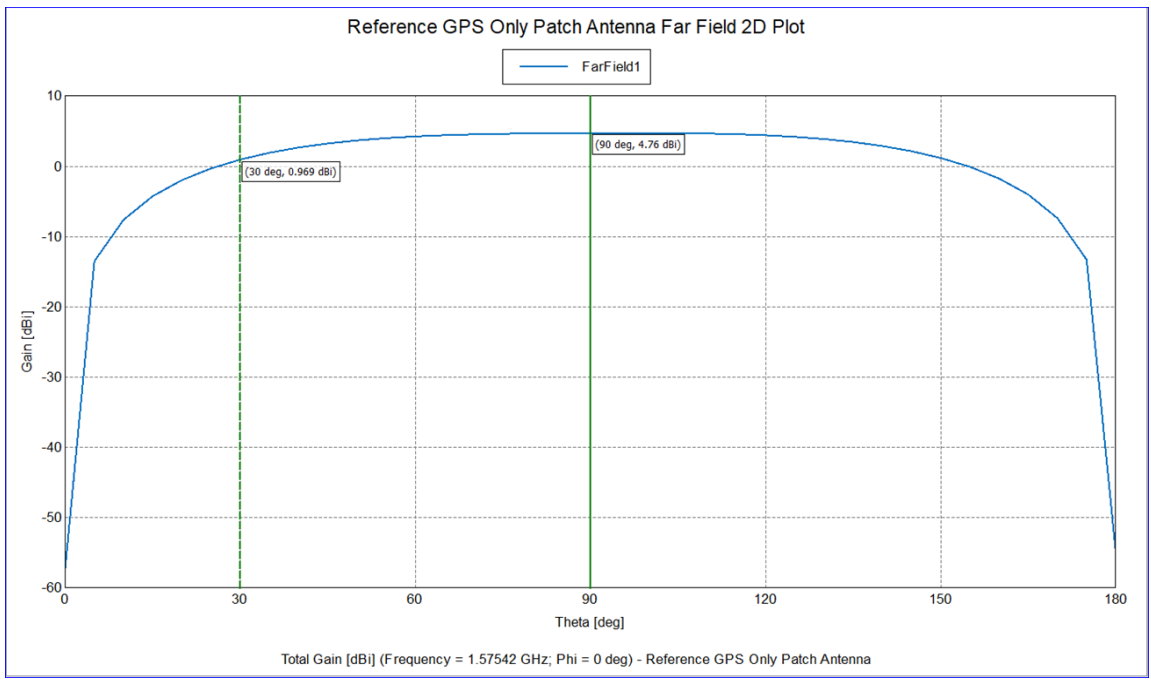

Figure 11. Patch antenna operating at 1.57542 GHz rectangular plot

Using Figure 11, the passive gain is approximately 3.791 dBi of the presented antenna and it can be determined by taking the difference in gain angle/delta between 30 and 90 degree angles in the 2D plot graphic.

Figure 11 shows the simulated passive gain of the proposed antenna and we selected substrate material to be Ceramic/Porcelain with dielectric constant/relative permittivity=5.5 mm to model, design, and simulate the presented RF GPS (L1) only passive patch antenna.

## 4.5. Device Under Test (DUT) Active GPS/GLONASS Patch Antenna

Modeling, design, and simulation based on the following design parameters listed in table 5 below.

Table 5. Design Parameters of Device Under Test (DUT) Active GPS/GLONASS Structure Patch Antenna

| Parameter | Value |
| --- | --- |
| Feed Length | 0.5 mm |
| Operating Frequency | 1.5925 GHz |
| Ground Plane Length | 95 mm |
| Ground Plane Width | 95 mm |
| Radiating Element Length | 12.25 mm |
| Radiating Element Width | 12.25 mm |
| Substrate Length | 24.7 mm |
| Substrate Width | 24.7 mm |
| Substrate Thickness | 4.5 mm |
| Substrate Dielectric Constant (Relative Permittivity) for Ceramic/Porcelain Material | 5.5 mm |

Figure 12 plot shows the Top View of the active GPS/GLONASS patch antenna geometry with a square/rectangular orange color ground plane, square purple color substrate with dielectric constant value of 5.5 mm and a square dark blue color radiating element where the model was created in CADFEKO. Figure 13, Figure 14, and Figure 15 depicts the Cross Section View, Front View, and Side View respectively of the active GPS/GLONASS patch antenna within the POSTFEKO environment.

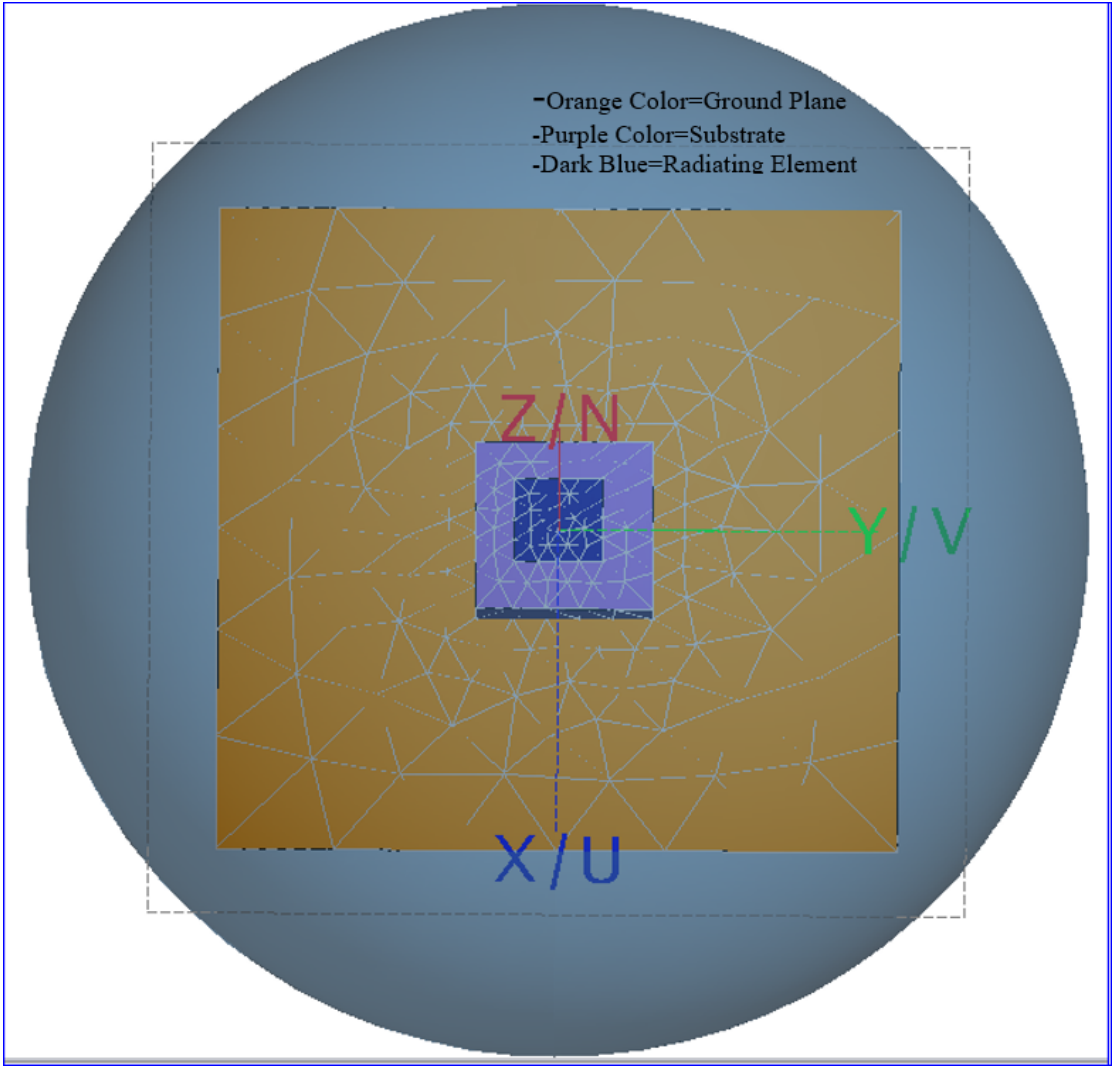

Figure 12. GPS/GLONASS Patch operating at 1.5925 GHz (Top View-CADFEKO v2020)

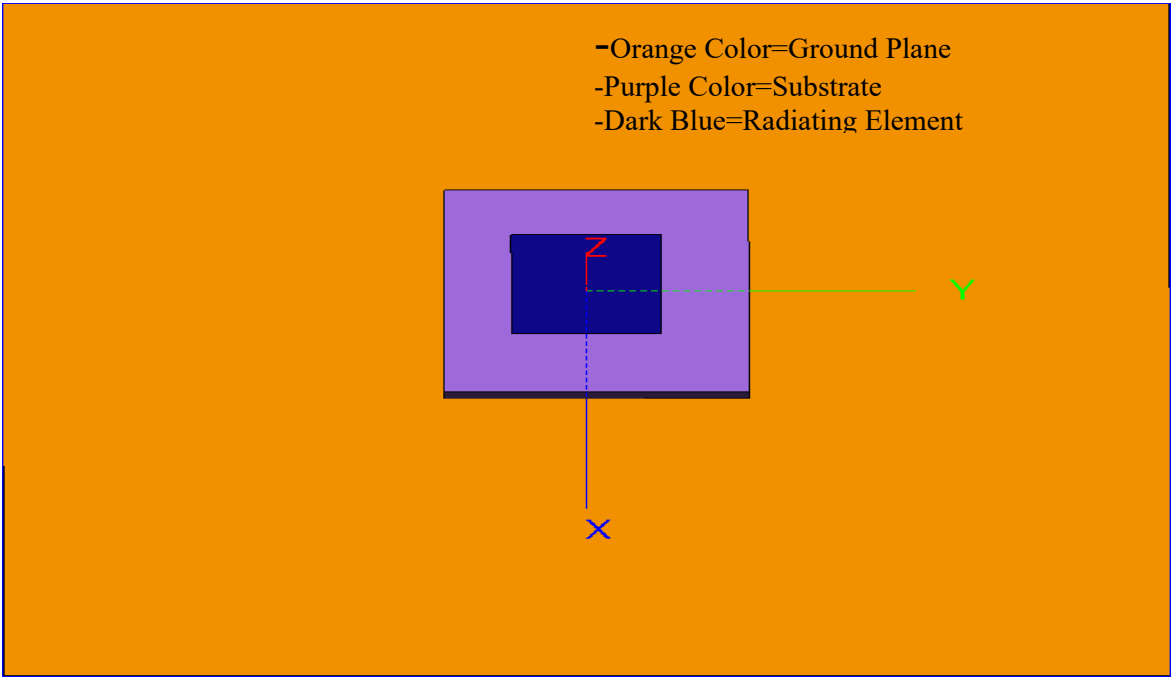

Figure 13.  GPS/GLONASS Patch operating at 1.5925 GHz (Cross Section-POSTFEKO v2020)

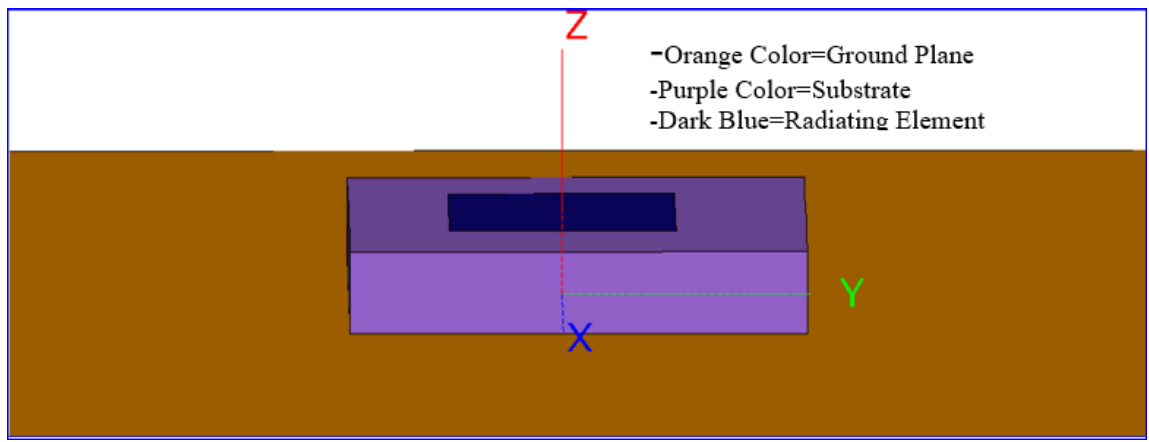

Figure 14.  GPS/GLONASS Patch operating at 1.5925 GHz (Front View)

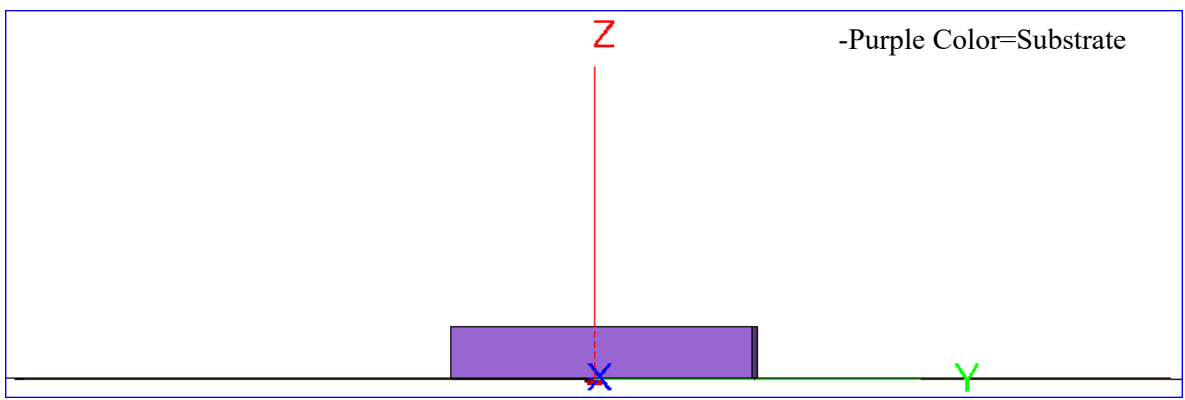

Figure 15.  GPS/GLONASS Patch operating at 1.5925 GHz (Side View)

**4.6. Simulated Far Field of GPS/GLONASS Structure Patch Antenna**

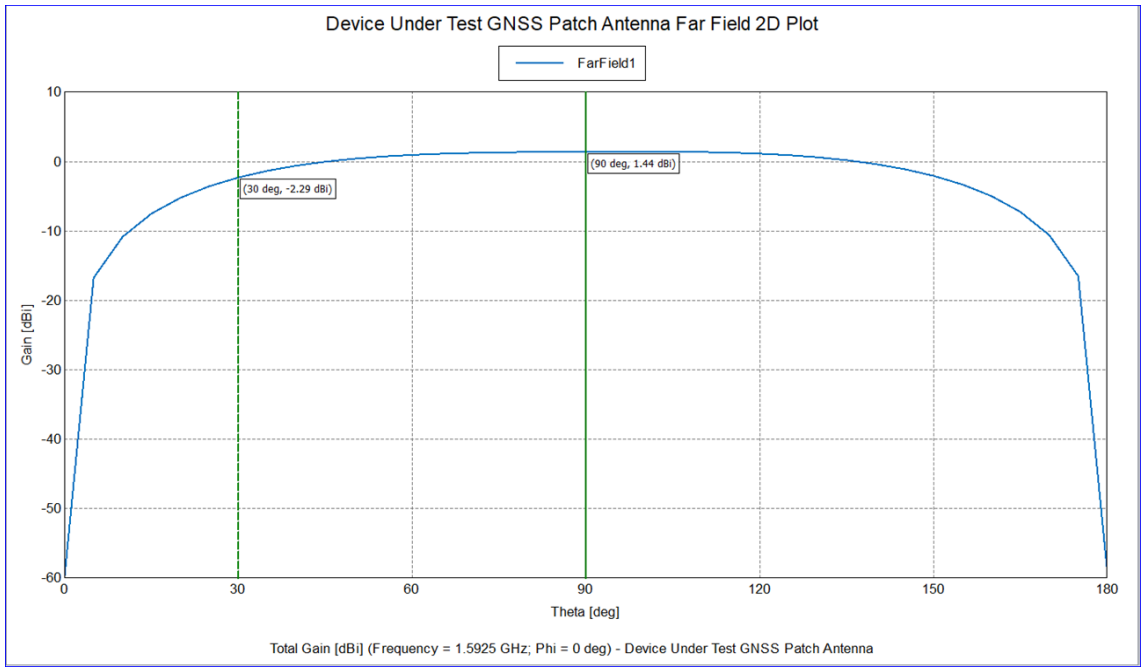

Figure 16.  Patch antenna passive gain at 1.5925 GHz rectangular plot

In the 2D model plot from Figure 16, we can see about 0.85 dBi passive gain, by taking the difference in gain between 30 and 90 degrees.

Figure 16 simulated the passive gain of the proposed antenna and we selected substrate material to be Ceramic/Porcelain with dielectric constant/relative permittivity=5.5 mm to model, design, and simulate the presented GPS/GLONASS patch antenna.

## 5. Conclusions

This paper describes the GPS (L1 1.57542 GHz) frequency) patch antenna performance and compares it to that of GPS/GLONASS patch antenna with ceramic/porcelain substrate material. These two antennas can be used in modern automotive applications. The models for each antenna were developed and then simulated on FEKO. The performance characteristic, such as passive gain in dBi were found, 3.791 dBi for GPS and   0.85 dBi for GPS/GLONASS delta between 30 and 90 degrees. The simulated results show an improved passive gain for the antenna. Thus, the proposed GPS/GLONASS will meet the needs of future automotive applications in robust way. Furthermore, other characteristics such as wide band width and efficiency are examined.

## 6. FUTURE WORK SUGGESTIONS

In future work this presented antenna can be modified, further studied and simulated in each of the following manners:

- The substrate material type can be changed from ceramic/porcelain to non-ceramic version/variant.

- The proposed antenna performance can be further improved by selecting a thick substrate whose relative permittivity is in the lower/smaller value than the presented dielectric constant of 5.5 mm.

- The mechanical dimensions of each antenna can be altered and simulated, in order to enhance antenna major performance parameters, such as Directivity, Impedance, Current and Polarization.
- The verification, validation and testing of the proposed RF GPS only patch and GPS/GLONASS patch antenna component can also be conducted at the system vehicle level, where each of the presented antenna can be installed and mounted on an optimal vehicle roof location area prior to the start of the testing and antenna performance parameters experimental measurement can be ascertained.
- The testing, assessment and evaluation of the presented RF GPS only (L1 1575.42 MHz frequency) and GPS/GLONASS patch antenna can be carried out in Anechoic Chamber and/or Indoor Antenna Range, in order to measure the basic antenna performance parameters and/or characteristics, such as Radiation Pattern, Radiation Efficiency, Directivity, Impedance, Polarization and Current draw and/or Antenna Surface Current.

## AUTHORS

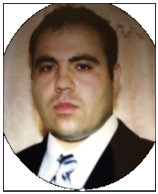

**Gholam Aghashirin** graduated from Ryerson University, Toronto, Ontario Canada with a B.Eng. in Electrical, Electronics and Communication Systems, earned his M.Sc. in Electrical and Computer Engineering from Oakland University, Rochester, Michigan, USA and he is currently a Ph.D. candidate in Electrical and Computer Engineering at Oakland University, Rochester, Michigan, USA. He has worked as an Engineer in advanced engineering projects, assignments in the automotive industries at various level of complexity and leadership roles in the field and space of Global Telematics, Automotive Radio Head Units, Navigation Systems, Instrument Clusters, Voice Recognition, Dialog, Hands-Free Systems, Electrical and Electronics ADAS and Automated Driving L2 and L3 Systems. His research interests include Electromagnetics, Location Technologies, antenna design, modeling, simulations at the component, vehicle level, antenna range and antenna experimental measurements.

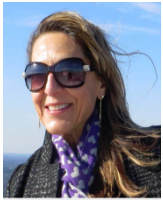

**Hoda S. Abdel-Aty-Zohdy** received the B.A.Sc. degree (with First Class Honors) in Electrical and Communications Engineering from Cairo University, the M.A.Sc and Ph.D. degrees in Electrical Engineering from the University of Waterloo, ON, Canada. Dr. AbdelAty-Zohdy is a Professor of Electrical and Computer Engineering, The John F. Dodge Chair Professor of Engineering, 2012-2014; Director of the Microelectronics & Bio-Inspired Systems Design Lab at Oakland University, Rochester, MI, USA. Her research and teaching focus on Circuits, Devices, VLSIC, H/W deep-learning, Electronic-Nose, and Bio-Inspired IC chips for high fidelity classifications. She organized, chaired, served on several conferences and committees for the IEEE/CASS and as Distinguished Lecturer 2004-2006,

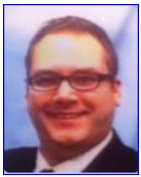

**Adam Timmons** received the Ph.D. degree in Materials Science from Dalhousie University, Halifax, Nova Scotia, Canada. Dr. Timmons is a Adjunct Professor within the Department of Mechanical Engineering at McMaster University, Hamilton, ON, Canada. He has many professional and academic appointments and holds a large number of patents.

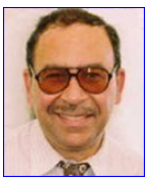

**Mohamed A. Zohdy** received the B.A.Sc degree in Electrical Engineering from University of Cairo, the M.A.Sc and Ph.D. (Medal) from the University of Waterloo, ON, Canada. Dr. Mohamed is a Professor of Electrical and Computer Engineering at Oakland University, Rochester, MI, USA. Professor Zohdy research focus is in the area of Advanced control and estimation, intelligent npattern information processing, neural, fuzzy, evolutionary systems, chaos control, smart simulation, hybrid systems.